# Exotic Transverse-Vortex Magnetic Configurations in CoNi Nanowires

Ingrid Marie Andersen[*,†], Luis Alfredo Rodríguez[‡,♠], Cristina Bran[§], Cécile Marcelot[†], Sébastien Joulie[†], Teresa Hungria[¶], Manuel Vazquez[§], Christophe Gatel[*,†], Etienne Snoeck[†]

[†]Centre d'Elaboration de Matériaux et d'Etudes Structurales – CNRS, 29 Jeanne Marvig, 31055 Toulouse, France
[♠]Centro de Excelencia en Nuevos Materiales, Universidad del Valle, A.A. 25360, Cali, Colombia
[‡]Department of Physics, Universidad del Valle, A.A. 25360 Cali, Colombia
[§]Instituto de Ciencia de Materiales de Madrid – CSIC, 28049 Madrid, Spain
[¶]Centre de Microcaractérisation Raimond CASTAING, Université de Toulouse, CNRS, UT3 – Paul Sabatier, INP, INSA, Espace Clément Ader, 3 rue Caroline Aigle, 31400 Toulouse, France

**Corresponding Authors**
*E-mail: mia.andersen@cemes.fr, christophe.gatel@cemes.fr.

**ABSTRACT** The magnetic configurations of cylindrical Co-rich CoNi nanowires have been quantitatively analyzed at the nanoscale by electron holography and correlated to local structural and chemical properties. The nanowires display grains of both face-centered cubic (fcc) and hexagonal close packed (hcp) crystal 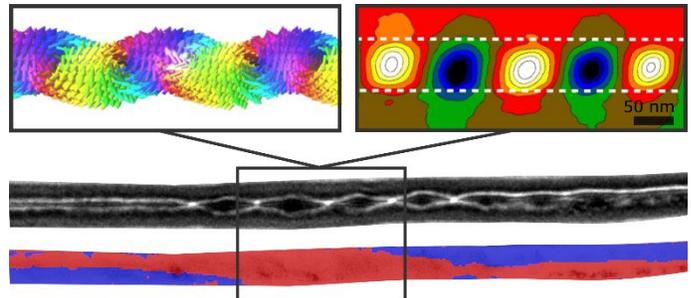 structures, with grain boundaries parallel to the nanowire axis direction. Electron holography evidences the existence of a complex exotic magnetic configuration characterized by two distinctly different types of magnetic configurations within a single nanowire: an array of periodical vortices separating small transverse domains in hcp rich regions with perpendicular easy axis orientation, and a mostly axial configuration parallel to the nanowire axis in regions with fcc grains. These vastly different domains are found to be caused by local variations in the chemical composition modifying the crystalline orientation and/or structure, which give rise to change in magnetic anisotropies. Micromagnetic simulations, including the structural properties that have been experimentally determined, allows for a deeper understanding of the complex magnetic states observed by electron holography.

One-dimensional magnetic nanostructures have for the past two decades been of increased interest for the development of future spintronic devices[1–4] motivated by concepts like Magnetic Race Track Memory[5] and the wish to manipulate magnetic domain walls (DW). Cylindrical nanowires (NWs) are particularly interesting candidates to reach this goal, much due to the fast DW motion induced by an external magnetic field or electric current, where theoretical studies have anticipated the absence of a Walker breakdown.[6,7] In addition, curvature effects have recently been proved to induce effects related to topology, chirality, and symmetry,[8] and unidirectional reversal process has been reported by engineering the geometry in multi-segmented nanowires.[9] However, to further technical developments in spintronics and a better control of DW motion, a thorough understanding of the fine structures of DWs in magnetic NWs, in which shape and crystal structure are contributing factors to the minimization of the system's magnetic energy,[10–13] is required.

The magnetic configurations in various Co-based cylindrical nanowires have previously been studied in several publications,[13–16] which have revealed a strong influence of the NW's structural properties,[17] and thus a dependence of their fabrication process.[10,18,19] For instance, it has been shown that monocrystalline hcp phase can be obtained in pure Co NWs with the c-axis engineered with nearly perpendicular orientation to the NW axis.[10,20] The corresponding uniaxial magnetocrystalline anisotropy is then strong enough to challenge the large shape anisotropy of NWs.[10,21,22] In CoNi-alloy NWs, the amount of Ni content can modify the crystallographic phase: while Co NWs with a low Ni content keep a hcp phase with a strong magnetic anisotropy and a resulting magnetic induction perpendicular to the NW axis, high Ni content generally leads to fcc crystal structure with a lower magnetocrystalline anisotropy[23,24] and a parallel magnetic induction. Tuning the $Co_xNi_{1-x}$ content then allows for adjusting the magnetic easy axis orientation from parallel (with cubic anisotropy) to perpendicular (with uniaxial anisotropy) relative to the nanowire axis,[10,25] making CoNi-alloy NWs interesting as potential building blocks for future devices.

However, previous publications have shown a coexistence of fcc and hcp crystal phases in single CoNi alloy NWs,[23,25] but there has been little or no report on how they mix and affect the magnetic configuration of the NW. Only statistical overviews of the structural and chemical information have been reported, as the crystal phase and composition in NWs are often determined by techniques like x-ray diffraction (XRD) and energy dispersive x-ray microanalysis (EDX) on assemblies or arrays of NWs.[10,18] To gain control over the magnetization states and reversal mechanisms, a precise analysis of the local structure of a single nanowire, combined with its magnetic configuration, is required. A quantitative study of both the structure and the magnetic properties obtained on the same area, and at the nano-scale, demands versatile and advanced techniques in combination with sufficient spatial resolution and sensitivity. Off-axis electron holography (EH) carried out in a transmission electron microscope (TEM) is an appropriate interferometric technique that allows for imaging of the magnetic configuration in nanostructures with nanometer resolution and high sensitivity.[26–30] EH offers quantitative information on the in-plane components of the magnetic induction inside the sample, as well as of the stray field surrounding it (see supplementary information S2). As EH is based in a TEM environment, it offers the advantage of combining other TEM techniques, like high-resolution TEM (HRTEM), selected area electron diffraction (SAED) and electron energy loss spectroscopy (EELS), and thus allows for studying the crystal structure and composition of the same area of the same NW with a high spatial resolution. In this work, we have combined EH with conventional TEM, Scanning TEM (STEM)-EELS and ASTAR™ measurements, a diffraction spot recognition technique based in the TEM (see supplementary information S3), to image and quantitatively determine the magnetic induction of small diameter cylindrical NWs of nominal composition $Co_{85}Ni_{15}$, *i.e.* NWs with a low Ni content, prepared by electrodeposition (see supplementary information S1), in relation to their local crystallographic structure and chemical composition. In addition, micromagnetic OOMMF simulations have been carried out for a deeper analysis of the local magnetic microstructure. This correlative study of TEM methods combined with micromagnetic simulations, allowed us to determine the key factors that govern the main parameters involved the magnetic configuration.

For the sake of consistency, all the results presented in this work are taken from the same area of a single nanowire, representative for the whole sample. Similar results were obtained from other NWs of the same batch.

**RESULTS AND DISCUSSION**

Figure 1a displays a TEM image of a representative $Co_{85}Ni_{15}$ NW of 70 nm diameter. The over-focused Lorentz image depicted in Figure 1b reveals an inhomogeneous magnetic configuration in the stippled region of the NW. The length of the enclosed area is 1.7 μm. EH analysis was carried out in this area and is shown in Figure 1c and 1d. Figure 1c displays the magnetic phase shift obtained at remanent state with a spatial resolution of 3 nm, and Figure 1d highlights the magnetic flux by applying a cosine function on the amplified version of the magnetic phase image in Figure 1c. A first analysis indicates that this NW displays a very complex magnetic configuration with four different magnetic regions, marked C1 to C4. The magnetic configuration in regions C1 and C4 in Figure 1c, presents isophase lines oriented parallel to the nanowire axis (x-axis as indicated by white arrows), evidencing an in-plane magnetic flux running along the NW axis and therefore a net component of the magnetization parallel to the NW axis. In contrast, the second prominent configuration of magnetic isophase lines, marked C2 in Figure 1c, displays a chain of circular patterns, aligned periodically in the middle of the nanowire. These curling isophase lines separate regions with magnetic flux pointing in opposite directions (marked by white arrows). This magnetic pattern corresponds to an antiparallel domain-like region whose magnetization is oriented perpendicular to the nanowire axis (marked by arrows along y-axis) with alternating opposite direction. Such a configuration is expected in NWs with magnetization easy axis oriented perpendicular to the nanowire axis as a consequence of a perpendicular magnetocrystalline anisotropy strong enough to counterbalance the NW shape anisotropy, which tends to align the magnetization parallel to its axis. The antiparallel coupling between adjacent domains allows for minimizing the dipolar energy. Lastly, the region marked C3 in Figure 1c bears resemblance to region C2, but with the isophase lines slightly oriented towards the nanowire axis, when compared to the C2 region. C3 region thus seems to be an intermediate state between C4/C1 and C2 regions. Our results thus demonstrate that three distinctly different magnetic domains appear within a limited length of the NW.

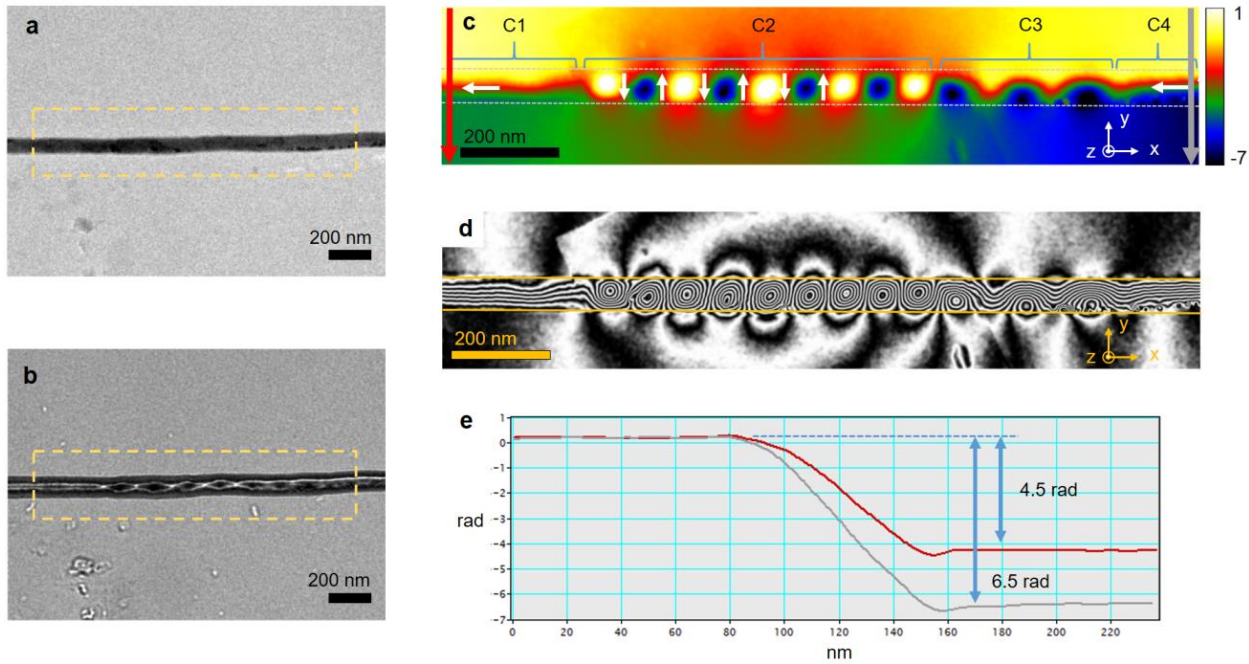

Figure 1: (**a**) BF-TEM image and (**b**) over-focused Lorentz image of sample NW. (**c**) Unwrapped magnetic phase shift image of stippled region marked in (**a**) and (**b**) with magnetic phase shift in radians represented with color scale, and (**d**) magnetic flux lines obtained from the cosine of (**c**) with amplification factor 4. (**e**) Plot of measured phase shift of cross sections in regions marked by red and gray arrows in (**c**).

We start a deeper analysis of the magnetic phase image by focusing our attention on the C1 and C4 areas, presenting a net component of the magnetization parallel to the NW axis. By measuring the magnetic phase shift across the NW in each of the C1-4 regions (marked in Figure 1c by red and grey arrows, and plotted in Figure 1e) and assuming a circular NW cross section with a uniform magnetization along the x-axis, we can extract the value of the in-plane magnetic induction parallel to the NW axis in the selected regions (see supplementary information S2). Using equation 6 from S2, we measure a magnetic induction of 0.7 T in C1 and 1 T in C4, while the expected saturation magnetization in $Co_{85}Ni_{15}$ alloy NWs, with all magnetic moments aligned parallel to the NW axis, is 1.6 T.[31] It then comes out of our experimental measurements that either the composition of the $Co_{85}Ni_{15}$ alloy is far from the one expected, leading to a strong decrease of the magnetization, or there is no region with a uniform magnetization parallel along the NW, even though the isophase lines in regions C1 and C4 indicate at it to be at least partly aligned along the axis. However, the hypothesis concerning a large deviation of the alloy composition is not valid: such magnetization values would correspond to a high Ni content of more than 70%. This is in contradiction with our TEM spectroscopy measurements as well as previous studies performed on similar NWs elaborated by the same procedure,[31] where a mean value of 15% Ni is expected. We thus conclude that the magnetization is not uniform and present a rotating component around the NW. In addition, the difference between measured values in regions C1 and C4 clearly indicates that even if the magnetic configuration looks similar in these regions, their total magnetic induction components parallel to the NW axis are different, and lower in the former compared to the latter.

This is qualitatively evidenced in Figure 1d, where the magnetic flux line density is much higher in the C4 region compared to C1.

Like all other TEM methods, EH is a projection technique: the phase shift of the electron beam related to the in-plane magnetic induction components, is integrated along the beam path. In addition, EH is only sensitive to the components of the magnetic induction perpendicular to the beam path, *i.e.* in-plane magnetic components (see Supplementary Information S2 for details). In case of a rotation of the magnetization around the NW axis, no integrated in-plane magnetic induction would be detected by EH as each magnetic induction component would have an equal, but opposite counterpart or will contain an out-of-plane component. As a result, the integration along the beam line would give zero phase shift, and only the remaining component along the NW axis would contribute to the phase shift. Accordingly, the magnetic configuration and the low phase shift value measured by EH in regions C1 and C4, indicate that a part of the magnetic induction presents a rotation around the NW axis. Two magnetic configurations, combining both rotating and parallel magnetic components, could in principle be considered: either a vortex state or a partial magnetic curling state around the NW axis. The possibility of vortex state is disregarded as it should give rise to inhomogeneous magnetization. The magnetic curling state has been reported by Ruiz-Gómez *et al.*[32] and is described by a magnetization at the core of the NW pointing in the direction of the nanowire axis, surrounded by a helical magnetic configuration. As previously stated, we observe a change in measured induction between regions C1 and C4 (Figure 1e) that could arise from a difference in size of the core of such a vortex state as discussed by Bran *et al.* and Ivanov *et al.*[31,33] or from a change in rotation angle of the curling with respect to the nanowire axis. The distribution of the magnetic shift and deeper analysis suggest that it is the curling state that occur in our system, where the highest phase shift value in region C4 comes from a smaller curling angle, *i.e.* a more elongated rotation along the nanowire axis, and thus a stronger measured magnetic induction.

Moving on, region C2 in Figure 1c bears evidence of antiparallel magnetic domain-like regions oriented perpendicular to the nanowire axis (along the y-axis), separated by states depicting a circular shape. These antiparallel domain-like regions are separated by a series of vortex states where the magnetization direction of the vortex core is oriented perpendicular to the nanowire axis (along the z-axis) with alternating chirality (rotation of the in-plane magnetization) or polarity (magnetic core orientation). Note the surprising exotic magnetic configuration where the transverse domain-like regions are significantly smaller in size than the vortex states. Finally, the C3 segment in Figure 1c looks like a transition region between the two already presented magnetic configurations (perpendicular vortex chain, and curling state) in C2 and C4: from C2 the vortex states disappear and the magnetic flux parallel to the NW axis increases as we move towards C4.

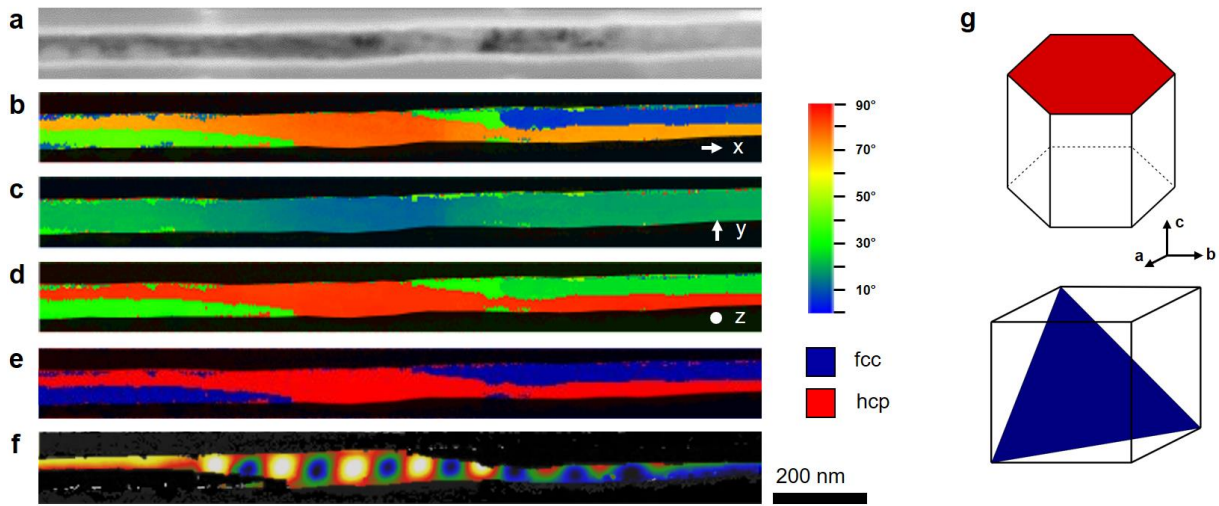

Figure 2: Experimental results of structural analysis of nanowire region. (**a**) Virtual bright field image of scanned nanowire area. (**b**)-(**d**) Crystal orientation map showing orientation of [111] and [0001] axis of respectively fcc grains and hcp grains by color code after angular displacement with respect to (**b**) x (**c**) y and (**d**) z axis. (**d**) Virtual crystal phase map of region in (**a**), where blue represent highest match for fcc phase, and red for hcp. (**f**) Superimposed magnetic phase shift image and virtual crystal phase map of the same nanowire region where only the hcp phase regions are visible. (**g**) The hcp and fcc facets from the orientation color code maps in (**b**)-(**d**)

To understand why this nanowire presents this complex magnetic configuration, we performed structural and chemical analysis on the same nanowire area, as shown in Figure 2a-e. ASTAR™ measurements were carried out to determine the local crystalline structure within the NW by retrieving precession diffraction patterns of the nanowire region (Figure 2a show mapped area), which were then analyzed by comparison to a database of calculated diffraction patterns using the automatic pattern recognition software of ASTAR™ NanoMEGAS[34,35] (See supplementary information S3 for more details). The results, as seen in Figure 2b-e, show that the region generally consist of three crystal grains: two fcc and one hcp crystal phase grains were found to coexist in the nanowire segment. The appearance of two phases has already been documented in CoNi alloyed NWs in literature.[23,25,31] Contrary to previous results where different phases were observed in different regions of the NW, we here see that the grain boundaries between the fcc and hcp phases are running almost parallel to the NW axis (Figure 2e). Thus, the two phases are coexisting within the same cross-section in regions of the NW but not with the same proportion along the NW. ASTAR™ experiments (Figure 2c) also indicate that the grains' close-packed directions $[111]_{fcc}$ and $[0001]_{hcp}$ are close to perfectly aligned with respect to the y-axis (perpendicular to the nanowire axis), making a transition from one crystal phase to the other possible by stacking fault. Figure 2c-d also show that the c-axis of the hcp structure is on average oriented 78° relative to the nanowire axis, *i.e.* almost perpendicular to it.

By overlapping the magnetic phase shift image onto the same region of the crystal phase map extracted from ASTAR™ (Figure 2f), we observe that the chain of vortex states (regions C2) is located on the area which consist mainly of hcp phase, indicating a direct correlation between the perpendicular magnetic configuration and the hcp structure with the [0001] direction (c-axis) at 78° relative to the NW axis (Figure 2c). It is well known that fcc and hcp crystal phases have different types magnetocrystalline anisotropy (uniaxial for hcp phase, and cubic for fcc phase), with different values of the anisotropy constant,[24,36] and consequently different easy axis directions. The local change of crystal phase is therefore assumed to be the origin of changes in magnetic configuration, and it is likely the magnetocrystalline anisotropy of the hcp phase that favors the magnetization alignment along the c-axis, as it is strong enough to challenge the shape anisotropy of the NW, which in turn promotes a stronger transverse magnetic component.[37]

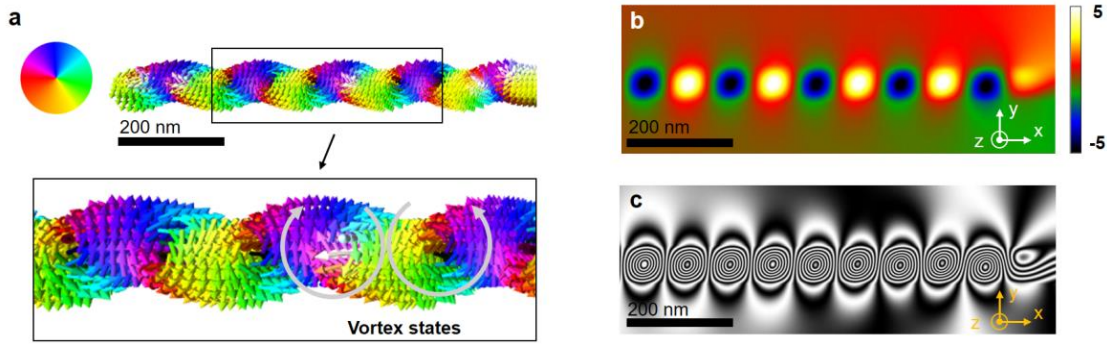

Figure 3: Results from micromagnetic simulation. (**a**) 3D representation of the simulation showing the magnetic vectors with their x-y-direction indicated by the color wheel. Zoomed images of the vortex state displayed below in (**a**). (**b**) Magnetic phase image and (**c**) magnetic flux lines obtained from the simulation.

To gain a deeper understanding of the complex magnetic configuration observed by EH, we performed a static micromagnetic simulation of the remnant state by the use of OOMMF code.[38] We used the magnetic parameters for a $Co_{85}Ni_{15}$ monocrystalline hcp phase structure for the hcp grain,[31] with saturation magnetization $M_s = 1273$ kAm$^{-1}$,[31] a value of the exchange constant $A = 26 \times 10^{-12}$ Jm$^{-1}$, and using an uniaxial magnetocrystalline anisotropy value of $K_1 = 350$ kJm$^{-3}$ oriented at 78° relative to the NW axis, as found by ASTAR™ for the hcp grain in the NW sample. Figure 3 shows the main results of the micromagnetic simulations, where Figure 3a shows the three-dimensional simulated magnetization at remnant state obtained from OOMMF calculations, where the x-y-plane of the magnetic vectors are represented by the colored directions of the color wheel. Figure 3b shows the simulated magnetic phase image calculated from the micromagnetic simulations, and the corresponding flux image is shown in Figure 3c We obtain a nice agreement between the experimental (region C2 in Figure 1c) and the simulated phase image (Figure 3b), allowing for a determination of the involved magnetic parameters through the OOMMF simulation.

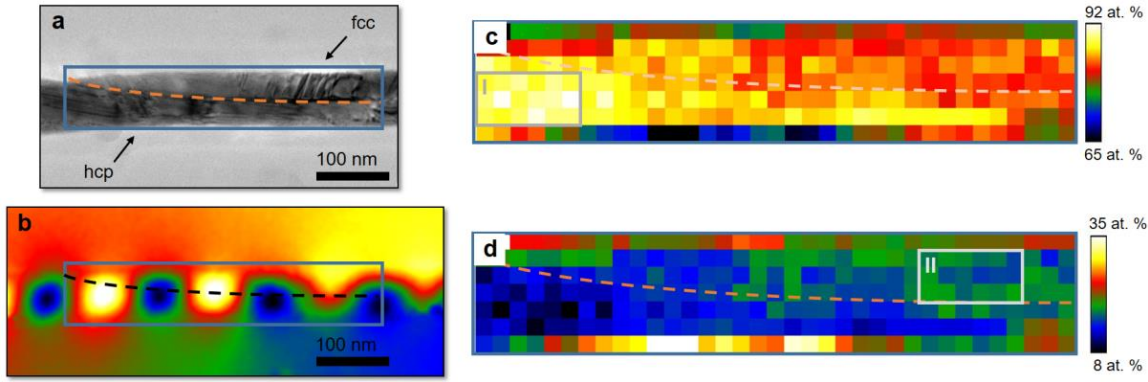

Figure 4: Experimental results of compositional analysis. (**a**) Bright field image. (**b**) Magnetic phase shift image of corresponding region. (**c**) and (**d**) show compositional maps of respectively Co and Ni composition from squared region in (**a**).

Local chemical analyses were performed by spectrum images in Scanning Transmission Electron Microscopy – Energy Electron Loss Spectroscopy (STEM-EELS) mode to elucidate the origin of the observed hcp/fcc structure change. We focused our STEM-EELS analysis around the hcp-fcc grain boundary, which is visible in the bright field TEM image in Figure 4a and corresponds to the magnetic configuration shown in Figure 4b. From spectrum imaging experiment carried out on the $L_{2,3}$ edge of Co and Ni, we have extracted Co and Ni content maps (Figure 4c and 4d) using multiple linear least squares (MLLS) fitting.[39] These maps reveal a difference in composition between the two grains; the region of hcp structure show a higher Co concentration (average of 89% Co and 11% Ni in square I in Figure 4c) as compared with the fcc structure (average of 80% Co and 20% Ni in square II in Figure 4d), which appears contain less Co. This agrees well with Co-Ni phase diagrams, where hcp phase appears in Co-rich [Co-Ni] alloys.[24] The lowest Co content in square I is 85.5%, while the highest Co content in square II is 82%. This point to a compositional threshold around 80-85% of Co to stabilize hcp phase, which is higher than the previously reported threshold of 70-75% Co[24,40] at room temperature for film or in the bulk. However, a more systematic and extensive study of this must be made to clearly conclude an accurate threshold, which partly depends on the elaboration process.

The chain of vortex states is located in the large grain of mainly hcp phase (C2), and the curling state seems to be located in regions with a coexistence of fcc and hcp phase grain (C1 and C4). Spectrum imaging and ASTAR™ experiments indicate that the magnetic transition region C3 is located almost along the hcp/fcc grain boundary, where both the structure and the composition is likely to change. The drastic changes in magnetic configuration appear to be caused by local changes in crystal structure induced by inhomogeneous composition of the NW. Even for small-to-moderate changes in composition (Detected Ni content of 8-25%, excluding edge measurements), results in a change in the crystal structure, and from this, a drastic difference in the local magnetic configuration.

## CONCLUSIONS

Quantitative and local studies of the magnetic, structural and chemical changes have been performed on cylindrical textured Co-rich CoNi nanowires. A complex and inhomogeneous magnetic configuration has been revealed, consisting of a periodical configuration of exotic antiparallel transverse domain-like regions separated by transvers-vortex states of alternating chirality and/or polarity at the hcp rich region. In turn, in the more fcc rich regions, mostly axial domains are observed. A transition between these two regions has also been identified. The experimental results have been compared to micromagnetic simulations, revealing that the vortex chain is formed inside hcp phase of the Co-rich CoNi alloy. Correlated local changes in composition and crystal structure have been highlighted as the origin of the different magnetic configurations: the vortex chain is a result of the hcp phase whose magnetic easy axis (c-axis) is oriented 78° relative to the NW axis (x-axis), and counterbalances the shape anisotropy. A curling state with a magnetic induction component oriented along the NW axis is appearing in regions with fcc phase grains where the easy axis and shape anisotropy act in the same way. A transition region between these two configurations is observed at the fcc/hcp grain boundary running almost parallel to the NW axis.

This study demonstrates the strong correlations between crystal structure and composition, and consequently on the magnetic configurations in CoNi NWs. The effect of local crystalline changes due to a slight variation of composition is found to drastically change the magnetic configuration in the NW, and can therefore not be neglected when attempting to explain and analyze the magnetic structure. A correlative microscopy investigation has been a great advantage in this work, as it offers magnetic, structural and spectroscopic analysis of the exact same area, since the same sample can be inspected by the different microscopic techniques.

## METHODS

Cylindrical CoNi nanowires were prepared by electrodeposition into self-assembles pores of anodic aluminum oxide templates. For a detailed description of the synthesis, see supplementary information S1. The NWs were drop casted onto a TEM carbon grid for the EH, STEM-EELS and ASTAR™ measurements.

Studies of the magnetic configuration of the CoNi NWs were performed by EH using a Hitachi HF-3300 (I2TEM-Toulouse) transmission electron microscope operated at 300kV. This microscope is equipped with a cold field emission gun, a spherical aberration corrector, and a double biprism setup, which makes it a dedicated microscope for interferometry experiments. The experiments were carried out in the normal stage of the I2TEM, at remanence state (with objective lens switched off), after an application of 1.8 T perpendicular to the sample, by switching the objective lens on and off again. The holograms were acquired by the use of a software for fringe position correction developed by C. Gatel,[41] with an exposure time of 160 seconds per hologram, an inter-fringe distance of 1.5 nm, allowing for a spatial resolution of 3 nm for the treated magnetic phase images. For a more detailed description of the electron holography technique, see supplementary information S2.

Studies of the structural configuration of the NWs were performed on a Philips CM20-FEG TEM at 200kV using the NanoMEGAS ASTAR™ system, a TEM based automatic crystal orientation and phase mapping technique.[35]

The data was acquired using precession electron diffraction with spot size ~ 1 nm, camera length 235 mm, mapped with step size of 4 nm. See supplementary information S3 for a description of the ASTAR™ system.

Studies of the compositional distribution of the NWs were performed using STEM-EELS at a JEOL ARM200F TEM operated at 200kV. This microscope is probe-corrected and equipped with a cold FEG. The data was acquired using GIF Quantum ER imaging filter and an energy resolution of 0.33 eV (measured at the FWHM of the zero loss peak) in dual EELS mode, acquiring low loss and high loss spectra at the same time. For the data treatment we used MLLS fitting algorithm in Gatan Microscopy Suite®'s Element Quantification tool in order to analyze the signal from the overlapping Co and Ni edges. See supplementary information S4 for a description of the MLLS fitting procedure.

Micromagnetic simulations were performed using OOMMF[38] with the following magnetic parameters: saturation magnetization, $M_{s-hco}$ = 1273 kAm$^{-1}$ [31]; exchange constant, $A_{hcp}$ = 26 x 10$^{-12}$ Jm$^{-1}$, magnetocrystalline anisotropy, $K_{1-hcp}$ = 350 kJm$^{-3}$ (uniaxial anisotropy, easy axis oriented 78° relative to NW axis). A representative 3D shape of the NW, with diameter of 70 nm, was built by stacking magnetic unit cells of 5 x 5 x 5 nm$^3$.

## ASSOCIATED CONTENT

**Supporting Information**

Supporting Information accompanies this which gives detailed information about the synthesis of the sample, a basic description of electron holography and of ASTAR™, as well as a description of the EELS analysis to separate the Co and Ni edges.

       S1: Synthesis of nanowires

       S2: Electron Holography

       S3: ASTAR™

       S4: EELS analysis by MLLS fitting

This material is available free of charge *via* the Internet at http://pubs.acs.org.

**Financial Interest**

The authors declare no competing financial interest.

## AUTHOR CONTRIBUTIONS

I.M.A. performed the EH experiments, and the associated data treatment. L.A.R. performed the micromagnetic simulations, and assisted I.M.A in EH experiments. T.H. performed the EELS measurements supervised by C.M. and I.M.A, and C.M. performed the data analysis. S.J. performed the ASTAR™ measurements supervised by I.M.A. C.B. fabricated the NWs. M. V. contributed in discussions and the development of the manuscript. C.G. and E.S. supervised this work and wrote the preliminary manuscript together with I.M.A.

**ACKNOWLEDGMENTS**

The research leading to these results has received founding from the European Union Horizon 2020 research and innovation program under grant agreement No. 823717 – ESTEEM3. The authors acknowledge the French National Research Agency under the "Investissement d'Avenir" program reference No. ANR-10-EQPX-38-01 and the "Conseil Regional Midi-Pyrénées" for financial support within the CPER program. This work was also supported by the French national project IODA (ANR-17-CE24-0047) and the international associated laboratory M$^2$OZART. This work has been partly supported by the Spanish Ministry of Economy, Industry and Competitiveness (MINECO) under project MAT2016-76824-C3-1-R and by the Regional Government of Madrid under project S2018/NMT-4321 NANOMAGCOST-CM.